\documentclass[aps,prl,twocolumn,groupedaddress]{revtex4}
\usepackage{graphicx}

\begin{document}


\title{Comment on "Indispensable Finite Time Correlations for Fokker-Planck
Equations from Time Series Data"}


\author{R. Friedrich$^1$
}

\author{Ch. Renner$^2$, M. Siefert$^2$, and J. Peinke$^2$
}
\affiliation{$^1$Westf\"alische Wilhelms-Universit\"at M\"unster
Institut f\"ur Theoretische Physik
D - 48149 M\"unster, Germany\\$^2$Carl-von-Ossietzky-Universit\"at Oldenburg,
FB - 8 Physics,
D - 21111 Oldenburg, Germany
}


\date{\today}

\begin{abstract}
\end{abstract}


\maketitle
Ragwitz and Kantz \cite{Kantz} propose a correction to a method for
the reconstruction of Fokker-Planck equations from time series data.
In  \cite{PRL,Siegert1,Siegert2,JFM} a method was presented which
directly applied the mathematical definitions of the drift $D^{(1)}$
and diffusion term $D^{(2)}$ \cite{Kolmogorov} for an estimate from
time series. Here different moments of conditional probability
densities (pdf) for finite step sizes $\Delta$ in the limit $\Delta
\rightarrow 0$ have to be estimated.  Ragwitz and Kantz state that
previous results have not been checked and that indispensable finite
time step $\Delta$ correction have to be employed for reliable
estimates of $D^{(2)}$. We want to add the following comments.  

Ragwitz and Kantz base their investigation on an estimate of the finite time
conditional probability in terms of a Gaussian,
eq. (7) of their paper. There is, however, no reason that for
finite $\Delta$ the conditional pdf is Gaussian.
The exact expressions for the conditional moments
up to the order $\Delta ^2$ can be unambigously derived from the
Fokker-Planck equation \cite{note}:
\begin{eqnarray}
\lefteqn{\langle x-x_0|x_0\rangle =\Delta D^{(1)}+}&&\\&+&\frac{1}{2}\Delta
^2[D^{(1)}(D^{(1)})' +D^{(2)}(D^{(1)})''] +\mathcal{O}(\Delta^3)
\nonumber\\
\lefteqn{\langle(x-x_0)^2|x_0\rangle= 2\Delta D^{(2)}+\Delta
^2[(D^{(1)})^2+}&&\\ &+&2D^{(2)}(D^{(1)})' +D^{(1)}(D^{(2)})'+D^{(2)}(D^{(2)})'']
+ \mathcal{O}(\Delta^3),\nonumber
\end{eqnarray}
For $\langle (x-x_0)^2|x_0\rangle$
the ansatz (7) of \cite{Kantz} neglects the last two terms which
are important for processes involving multiplicative noise. However,
intermittency of turbulence is related to a multiplicatice process. This 
remark especially applies to the wind data presented in
\cite{Kantz}. The validation of their method based on a Langevin
(equation (9)) only works since purely additive noise is considered.

For the investigation of turbulence
Ragwitz and Kantz claim to obtain remarkable correction, as shown
in their Fig. 6. Our approach in \cite{JFM} is based on an estimate 
of the diffusion term
using the limit $\Delta \rightarrow 0$ yielding a dependency which
can be approximated by a low order polynomial. In order to improve
on this estimate the coefficients of this polynomial have been varied
in such a way that the solution of the corresponding Fokker-Plank
equation for finite values of $\Delta $ yields an accurate 
representation of the measured one. In other words, in a
second step, we have performed a parametric estimation of the diffusion term.
In Figure 1 we present a case where a large correction to the
$\Delta \rightarrow 0$ estimation of $D^{(2)}$ had to be introduced 
(usually corrections are of the order of some percentages).
For finite values of $\Delta$ the estimated values
of  $D^{(2)}$ clearly differ from the limiting case $\Delta
\rightarrow 0$.
Estimations with different  "correction" terms for $D^{(2)}$ and for
finite $\Delta$ values may fake large corrections values.
Taking the limit $\Delta \rightarrow 0$ these deviations vanish
within the error.

The range of $\Delta$ which can be taken for the
estimate of $D^{(2)}$ must be choosen carefully in order to ensure that the
Markovian property holds, see \cite{JFM}. Since for each estimated value of
$D^{(2)}$  a finite number of data point is used, a
statistical error can be estimated for  $D^{(2)}$ (cf. Fig. 1).
These errors
naturally increase considerably for large values of x (compare Fig. 6
\cite{Kantz} and Fig 13 \cite{JFM}.)
%
%
\begin{figure}
\includegraphics[width=2in]{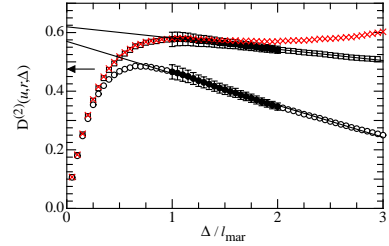}%
\caption{  $\Delta$ dependence of $D^{(2)}(u=\sigma_{\infty},r=L/2,\Delta)/\sigma_{\infty}^2$  for
different correction terms; squares without correction term, circles
with $((D^{(1)})^2)$, and
crosses with the correction term of \cite{Kantz}.  L denotes the integral length, $\sigma_{\infty}$ the rms of the velocity increments at large scales. Only for
$\Delta/l_{mar} > 1$ Markovian properties hold, and estimations of $D^{(2)}$
are senseful. The optimal value
of $D^{(2)}$ based on verifications
is indicated by an arrow, further detail see \cite{JFM}.
\label{}}
\end{figure}
To conclude, a deeper understanding of finite time correlations are of 
interest. The correct terms of higher orders in $\Delta$
\cite{note} may be used to improve on the estimation of drift and
diffusion terms. Up to now the best way for estimating
diffusion coefficients is to combine a nonparametric estimate
for $\Delta \rightarrow 0$ with a functional ansatz, i.e. a suitable
polynomial ansatz. Refining the estimates of the coefficients 
by parametric methods as for instance described by \cite{Timmer}
leads to improved results by a comparison 
of measured and calculated conditional pdfs at finite $\Delta$. 
\bibliography{basename of .bib file}

\begin{thebibliography}{10}
\bibitem{note} The Taylor expansion of the conditional
expectation of a function $f(x)$ is simply $\langle f(x(\Delta))|x\rangle=
\sum \frac{\Delta^n}{n!} (L^{\dagger})^n f(x)$, where $L^{\dagger}$ is the
adjoint Fokker-Planck operator.
\end{thebibliography}

\end{document}